# iDIGIT4L. Nuevos ecosistemas de digitalización y aprendizaje hombre-máquina para sistemas de fabricación industrial heredados

**Álvaro García García (Fundación Cidaut)**, Esteban Cañibano Álvarez (Fundación Cidaut)

## RESUMEN

La digitalización del ecosistema productivo hombre-máquina se ha convertido en una prioridad para las pequeñas y medianas empresas. Especialmente para hacer frente a los retos de la Industria 4.0 y a las habilidades digitales avanzadas en el puesto de trabajo. Desde el punto de vista de la investigación, la digitalización abre un escenario global de generación de oportunidades para el aprendizaje en sistemas de fabricación existentes. En particular, los entornos tradicionales deben lidiar con presiones competitivas para incorporar las nuevas tecnologías y adecuar las capacidades de los trabajadores. En este trabajo presentamos el proyecto iDIGIT4L, cuyo principal objetivo es la investigación y desarrollo de un ecosistema de digitalización donde intervienen personas y sistemas con el fin de transformar procesos industriales de forma inteligente y predictiva. Aporta una metodología de aprendizaje hombre-máquina en tres niveles que proporciona una interacción bidireccional y aumentada en un escenario de fabricación tradicional. Se apoya en la implementación de un gemelo digital integrado de forma no intrusiva, caracterizando una vieja fresadora industrial para el aprendizaje con modelos de conocimiento que cuentan con la experiencia de trabajadores expertos. Como resultado se han podido actualizar al mismo tiempo las funcionalidades del sistema industrial y las habilidades digitales de los trabajadores convirtiéndose en una parte más del gemelo digital.

**Palabras clave**: aprendizaje, retro-actualización, gemelo digital, Industria 4.0, realidad aumentada

## ABSTRACT

The digitization of the productive ecosystem related to human-machine interaction has become a priority for small and medium-sized enterprises. Particularly to face the challenges of Industry 4.0 and advanced digital skills in the workplace. From the research point of view, digitization opens a global scenario for the generation of opportunities for learning in existing manufacturing systems. Mainly, traditional environments must deal with competitive pressures to incorporate new technologies and adapt the skills of workers. In this paper is presented the iDIGIT4L project, which was envisaged to research and develop a digitization ecosystem where people and systems interact in order to transform industrial processes in an intelligent and predictive way. The project contributes with a three-tier human-machine learning methodology that provides augmented and bi-directional interaction in a traditional manufacturing scenario. It is based on the implementation of a non-intrusively integrated digital twin, characterizing an old industrial milling machine for learning through knowledge models supported by the experience of skilled workers. As a result, it has been possible to simultaneously update the functionalities of the industrial system and the digital skills of the workers, becoming an integral part of the digital twin.

# 1. INTRODUCCIÓN

Uno de los pilares que sustentan la economía europea es la industria. En particular, el sector de fabricación con aproximadamente 2,1 millones de empresas en Europa, tiene un papel clave junto a toda su cadena de valor. Atendiendo al cambio de paradigma que introduce la cuarta revolución industrial [1], las empresas, para seguir siendo competitivas, deben poder beneficiarse de las crecientes oportunidades digitales. Además, se da la circunstancia que del total del sector el 59% son Pequeñas y Medianas Empresas (PYMEs), obligadas a enfrentarse a una mayor carencia de recursos digitales y de trabajadores formados en nuevas tecnologías [2]. En el caso de España este problema está todavía más acentuado llegando a representar las PYMEs más del 99% del total.

Con el objetivo de mejorar la competitividad y contribuir a la reducción de la brecha digital, la Comisión Europea ha apostado por el Programa Europa Digital (2021-2027) para el apoyo a la transformación digital de la economía, el cual está dotado con un presupuesto de 9.200 millones de euros. La estrategia está destinada en buena parte a tecnologías habilitadoras. Entre ellas destacan las habilidades digitales avanzadas en el puesto de trabajo, especialmente para hacer frente a las necesidades de la Industria 4.0 [3]. Por lo tanto, la digitalización del ecosistema productivo hombre-máquina se ha convertido en una prioridad. Concretamente, la iniciativa I4MS (ICT Innovation for Manufacturing SMEs) hace énfasis en que, para avanzar hacia la industrialización y la adopción de la tecnología, especialmente en las PYMEs, es necesario innovar con metodologías que puedan testear su impacto sobre los procesos de fabricación [4]. El propósito es utilizar la información generada en los sistemas ciberfísicos para mejorar sustancialmente, entre otros, la capacitación del personal, la calidad del proceso, la optimización de recursos y la flexibilidad.

Desde el punto de vista de la investigación, la digitalización abre un escenario global de generación de oportunidades para el aprendizaje, donde entidades gubernamentales, investigadores, expertos en nuevas tecnologías y empresas ya están trabajando juntos en vías de aplicación de las tecnologías digitales en las fábricas. La necesidad de comprender el ciclo de vida y de cómo mejorar los procesos de fabricación asociados, proporciona nuevas formas de desarrollar métodos de generación de conocimiento hombre-máquina [5]. En esta línea, los sistemas de fabricación existentes deben lidiar con presiones competitivas e incorporar nuevas tecnologías y perfiles formados que permitan el desarrollo de aplicaciones a través de plataformas conectadas. Muchos de estos sistemas de fabricación son heredados y heterogéneos, por lo que deberán retro-actualizarse [6] para incorporar métodos de comunicación hombre-máquina que permitan estar conectados digitalmente e intercambiar datos e información tanto en la planta como más allá de la fábrica [7].

En este trabajo presentamos el proyecto iDIGIT4L, cuyo principal objetivo es la investigación y desarrollo de un ecosistema de digitalización donde intervienen personas y sistemas con el fin de transformar procesos industriales de forma inteligente y predictiva. Concretamente, en él se aborda el aprendizaje hombre-máquina para la mejora de un proceso de fabricación por mecanizado industrial a través de la virtualización de activos, independientemente de su nivel de digitalización. Para ello, implementa una solución de gemelo digital apoyada en la retro-actualización, que aplica una metodología de tres niveles para la convergencia de los sistemas industriales. Integra indicadores de estado y operación, procedentes de diversas fuentes heterogéneas con el uso de sensores no intrusivos conectados sobre una plataforma aumentada. Con el conocimiento adaptativo de las operaciones de mecanizado industrial, la plataforma desarrollada proporciona un método de digitalización flexible del entorno industrial propuesto. La unión de todos los componentes e interfaces hombre-máquina del gemelo digital, proporciona además la visualización en planta con realidad aumentada. De esta forma, la tecnología se pone al alcance de los trabajadores convirtiéndose en una parte más del gemelo digital para entender qué está pasando. Para demostrar la aplicación de la metodología se presenta un caso de uso de aplicación de retro-actualización no intrusiva y modelado con un gemelo digital sobre una fresadora industrial de tres ejes con una antigüedad de más de 25 años. En este entorno de mecanizado tradicional se han utilizado estrategias de aprendizaje hombre-máquina para la detección de anomalías durante el proceso de arranque de la fresadora, lo que permite anticipar la toma de decisiones durante tareas de mantenimiento y fabricación.

El resto de secciones se organiza como sigue: la Sección 2 presenta una revisión de la literatura. La Sección 3 describe la metodología de aprendizaje hombre-máquina basada en tres niveles. La Sección 4 presenta el caso de uso aplicado sobre la interacción bidireccional hombre-máquina, de forma no intrusiva, orientada a la detección de patrones de actividad en una vieja fresadora industrial. Finalmente, la Sección 5 presenta las conclusiones alcanzadas.

## 2. REVISIÓN DE LA LITERATURA

Una de las realidades actuales en las fábricas europeas es la falta de expertos formados y personal capacitado para la aplicación de las tecnologías de la información de nueva generación, que además se une a las barreras existentes para la digitalización y la fabricación inteligente [8]. En el caso de las PYMEs, la integración de antiguos sistemas de fabricación heredados, supone la introducción de retos adicionales para abordar la convergencia físico-digital [9]. Es un hecho que se necesitan nuevas habilidades digitales y herramientas de gestión del conocimiento para que la colaboración hombre-máquina sea efectiva a través de la integración de los activos industriales [10] y la interacción de los trabajadores en entornos tradicionales [11]; permitiendo procesos industriales conectados que promuevan el concepto de la Industria 4.0 [12]. De acuerdo a este enfoque, que involucra tanto a los trabajadores como a los sistemas industriales para la transformación digital de la industria manufacturera [13], el aspecto clave es cómo los sistemas industriales más antiguos se vuelven conectados. Todo ello, mientras se extrae la información y se genera el conocimiento para que los trabajadores se adapten de forma rápida y adecuada [14] en un nuevo contexto adaptativo [15].

El concepto de retro-actualización, dentro de los entornos de fabricación tradicionales, proporciona una vía para la conexión de máquinas heredadas mediante las tecnologías habilitadoras. En general, el proceso de retro-actualización proporciona una mejora de las máquinas añadiendo nuevas características técnicas, así como componentes electrónicos que se incorporan como parte del equipamiento (por ejemplo, dispositivos Human Machine Interface o HMI, sensores y actuadores) [6]. De forma adicional, la instalación de protocolos de comunicación, sistemas de captura de datos y rutinas de control, complementan las funcionalidades de conectividad [16]. En la literatura podemos encontrar casos de uso experimental en PYMEs que abordan la retro-actualización de una fresadora de Control Numérico por Computadora (CNC) utilizando metodologías basadas en conceptos de la Industria 4.0 [17]. En ese sentido, se ha abordado a nivel de laboratorio la demostración de que un sistema de fabricación tradicional puede ser retro-actualizado de forma no intrusiva a través de la implementación de un framework estandarizado acorde a la Industria 4.0 [18]. De esta forma, la utilización de tecnologías digitales y sensores permite la integración de datos de diferentes orígenes del proceso de fabricación a través de métodos de retro-actualización no intrusivos para la gestión de la monitorización de condiciones de una máquina [19]. En particular, la retro-actualización se puede aplicar para la detección del desgaste de herramientas en una fresadora CNC a través de un acelerómetro y modelos de clasificación de patrones [20]. Este tipo de estrategias permiten añadir al equipamiento heredado un registro de datos históricos, herramientas de análisis y un ecosistema de gemelo digital con una buena relación coste/eficiencia [21]. No solo esto, sino que además el potencial de la convergencia físico-digital [7] ha completado el círculo de la interacción entre sistemas y trabajadores [22]. La Industria 4.0 ofrece oportunidades para la implementación de ecosistemas de aprendizaje basados en el gemelo digital tanto en entornos industriales como académicos [23]. Sin embargo, frente a ese objetivo la industria se enfrenta a los retos de proporcionar capacidades y nuevas infraestructuras digitales, mientras que la academia se enfrenta a los retos de investigación en nuevas tecnologías para proporcionar programas formativos y expertos que den respuesta a las emergentes competencias digitales e interdisciplinares de los trabajadores [24]. La aparición de aplicaciones de aprendizaje basadas en el gemelo digital, presenta un nuevo escenario virtual que proporciona una capa adicional de conocimiento a los ecosistemas de fabricación. De este modo, la conectividad del gemelo digital permite diferentes modos de colaboración entre trabajadores y sistemas productivos automatizados. Además, la utilización de datos distribuidos para el aprendizaje ofrece la oportunidad de modelar las múltiples interacciones entre procesos [25], sistemas [26] y las competencias de los trabajadores [27].

Como tendencia a futuro en la industria, las fábricas de producción se apoyarán cada vez más sobre múltiples gemelos digitales para representar su sistema de producción completo [28]. Al promover las áreas de investigación abiertas, basadas en la aplicación de gemelos digitales, pueden surgir nuevos enfoques para transformar los métodos de producción y control existentes hacia interfaces ciberfísicas y modelos de apoyo a la toma de decisiones automatizadas. Tal es el caso de la interacción entre los trabajadores y el entorno de producción para el desarrollo de capacidades 4.0, lo que permite que el gemelo digital ofrezca un enfoque adaptativo al contexto de trabajo para apoyar la toma de decisiones y el aprendizaje [29].

## 3. METODOLOGÍA

Para la construcción de un ecosistema de digitalización y aprendizaje compuesto por activos que intervienen en el proceso de fabricación, se necesita una convergencia físico-digital de todos los actores implicados: sistemas, personas y procesos. En este trabajo se propone una metodología adaptativa y no intrusiva basada en el concepto de gemelo digital con tres niveles para la generación de conocimiento: (i) interacción, (ii) entendimiento y (iii) aprendizaje, como se muestra en la **Figura 1**. Además, este ecosistema se apoya en una plataforma multi-capa con herramientas de digitalización para la retro-actualización de sistemas industriales heredados y soluciones en la nube que ofrecen interfaces hombre-máquina.

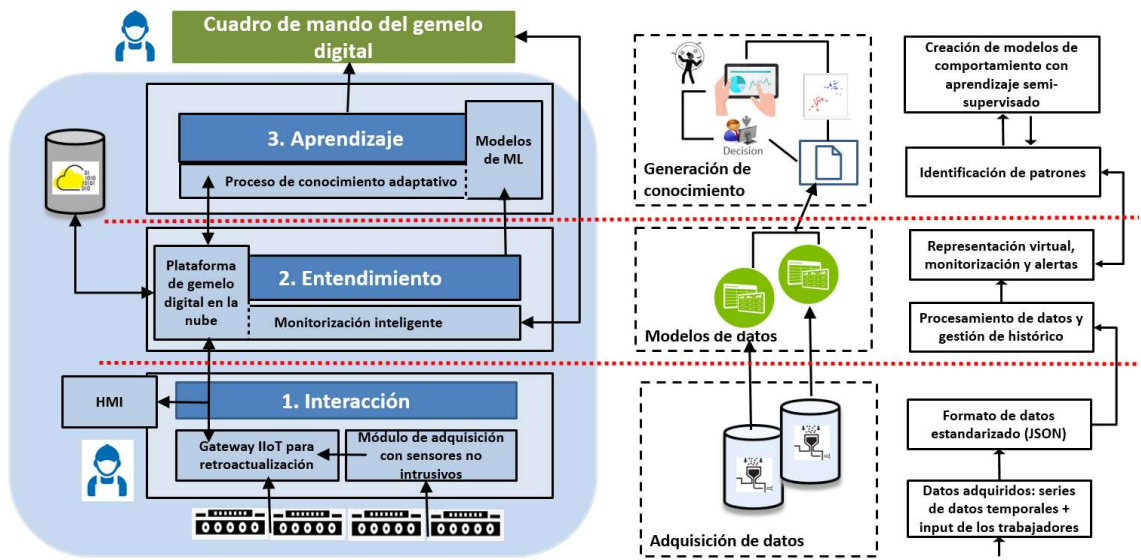

**Figura 1. Metodología de tres niveles propuesta para generar conocimiento a partir del concepto de gemelo digital**

El primer nivel de la metodología se encarga de la **interacción** a nivel ciberfísico. Para ello se han implementado interfaces de hardware y software estandarizadas sobre un Gateway industrial siguiendo una estrategia no intrusiva de retro-actualización. El objetivo es incorporar una capa modular de comunicación hombre-máquina sin hacer cambios en los sistemas de fabricación existentes. El Gateway, configurado de forma portable, proporciona un módulo de adquisición de datos con un rango extenso de señales de entrada/salida (por ejemplo: temperatura, vibración, consumo de energía, etc.) y los protocolos de comunicación necesarios para enviar los datos a la plataforma de gemelo digital en la nube. Los atributos de los diferentes puntos de medida, tales como acelerómetros, detectores de temperatura, analizadores de fase de corriente, etc. se dan de alta como objetos y se configuran de acuerdo a sus propiedades físicas (modo de operación estático/dinamico, rango de medida, unidades, etc.). Las lecturas registradas se guardan en el Gateway industrial utilizando una estructura de datos que contiene el sello de tiempo, ofreciéndose en formato estándar JSON a través de una interfaz API Web que permite la extracción de series de datos temporales.

El segundo nivel de la metodología proporciona modelos sobre los datos registrados y su visualización para el **entendimiento** de las distintas formas de interacción entre trabajadores, sistemas y procesos. Para ello, una plataforma en la nube conectada al Gateway industrial, gestiona y procesa los flujos de datos con la representación virtual de los activos, las propiedades y los indicadores asociados. Esta información se monitoriza sobre interfaces gráficas de usuario como dispositivos HMI. El despliegue de sistemas interactivos donde intervienen sensores y trabajadores a través de herramientas HMI, amplía la oportunidad de entender la interacción hombre-máquina con modelos de gemelo digital y generar conocimiento de los sistemas y los procesos. A través de un cuadro de mando en la nube se configuran y etiquetan todos los objetos correspondientes a los indicadores de proceso sensorizados. Además, los registros se modelan sobre gráficos y tablas de datos con umbrales para su correcta visualización e interpretación. El objetivo es lograr un entorno de aprendizaje altamente eficiente para los trabajadores que interactúan con procesos en tiempo real mejorando sus habilidades y atendiendo de forma temprana alertas e incidencias con estrategias de mantenimiento predictivo.

El tercer nivel de la metodología gestiona la generación de conocimiento a través de un proceso de **aprendizaje** adaptativo que caracteriza el proceso de fabricación. Toda la experiencia recopilada en los niveles anteriores, a través de los registros de datos correspondientes a los patrones de operación identificados durante la interacción hombre-máquina (arranque, puesta en marcha, funcionamiento, incidencias, parada, etc.), se combina con algoritmos de machine learning a través de estrategias de aprendizaje semi-supervisado como se describe en la **Figura 2**.

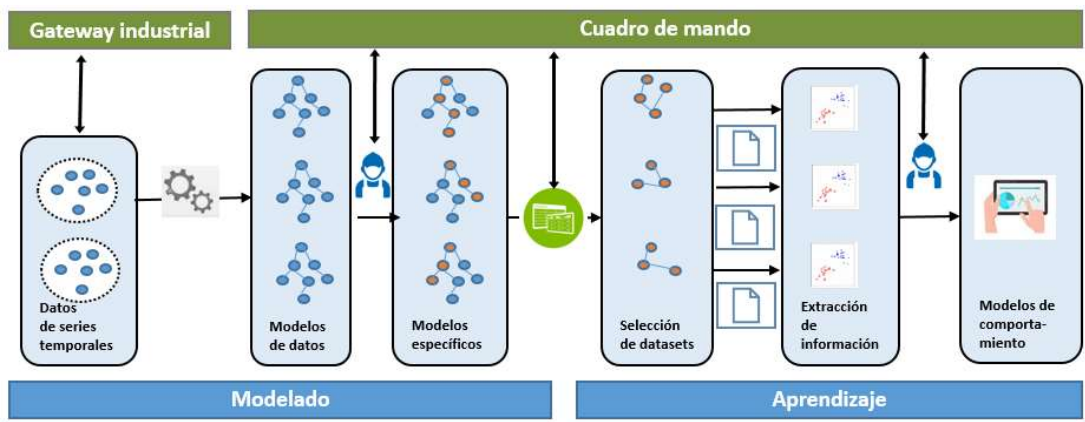

**Figura 2. Descripción del proceso secuencial de generación de conocimiento.**

Con la ayuda de personal experto de proceso se procede a clasificar las propiedades específicas de funcionamiento del sistema que se pretenden modelar. Entre ellas se encuentran ciclos o fases de trabajo concretas del sistema de fabricación, que se exportan en formato tabla. Se seleccionan los conjuntos de datos que incluyen las medidas físicas de acuerdo a la clasificación realizada por fases. A continuación, se procesan los datos para que tengan una escala común y se normalizan en el rango numérico de 0 a 1 para facilitar la comparación de resultados. Sobre las diferentes muestras de datos clasificadas se realiza la extracción de información oculta con herramientas de código abierto y algoritmos de aprendizaje automático basados en Python 3.x, para detectar el estado de salud del sistema analizado. Los patrones obtenidos se etiquetan por un operador experto para construir los modelos de comportamiento en base a las relaciones aprendidas de forma automática sobre los conjuntos de datos que caracterizan la operación de la máquina. De esta manera las lecciones aprendidas se convierten en modelos de referencia para dar apoyo a la toma de decisiones sobre las tareas de fabricación y mantenimiento basado en la condición del activo, a la vez que retroalimentan el proceso de aprendizaje sobre interfaces HMI, aplicaciones móviles de realidad aumentada con tabletas, e indicadores físicos publicados sobre el cuadro de mando del gemelo digital.

## 4. CASO DE USO

### 4.1 DESCRIPCIÓN

Con objeto de detallar y evaluar la aplicabilidad del proceso de generación de conocimiento propuesto se ha utilizado una fresadora Nicolás Correa CF-20 provista de un panel CNC Heidenhain TNC-407 que está ubicada en las instalaciones del Centro de Investigación y Desarrollo Fundación Cidaut (ver **Figura 3**). Se trata de un caso de uso de fabricación tradicional basado en una vieja fresadora CNC no digitalizada, donde los procedimientos de operación son programados manualmente por un trabajador del taller de mecanizado a partir de diseños CAD en un flujo de proceso no conectado. Esta fresadora es una máquina herramienta que normalmente se usa para diferentes trabajos de mecanizado de piezas de material sólido con una herramienta de corte giratorio. La fresadora NC CF-20 no cuenta con información histórica registrada y monitorizada durante su ciclo de vida, más allá de las estrategias de mantenimiento preventivo o correctivo. El operador de la máquina es experto en su manejo y se encarga del encendido y puesta en marcha al inicio de la jornada laboral, procediendo a su apagado al final de la misma. Con la situación de partida, sin información del proceso de fabricación, no es posible tomar decisiones de forma anticipada de acuerdo a mejoras del proceso o condicionadas por el estado de funcionamiento.

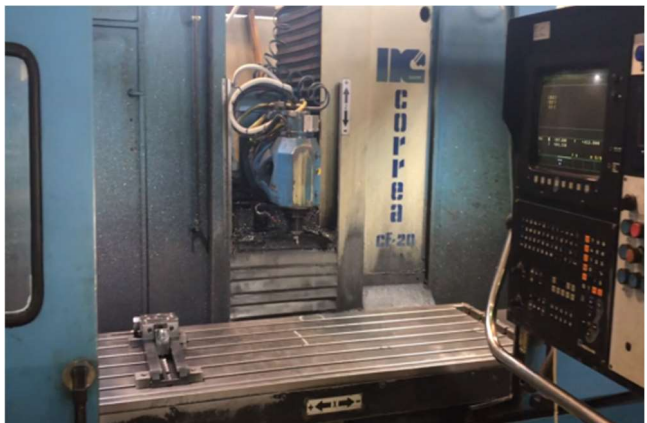

**Figura 3. Fresadora Nicolás Correa CF-20 en Cidaut.**

### 4.2 IMPLEMENTACIÓN

Para abordar el caso de uso del proyecto iDIGIT4L se desarrollaron de forma modular los tres niveles propuestos en la Sección 3, aplicándose sobre la fresadora NC CF-20 con la ayuda de un operador experto. La adaptación de las herramientas digitales al espacio de trabajo en el taller de fabricación se utiliza para habilitar una interacción bidireccional hombre-máquina, de forma no intrusiva, orientada a la detección de patrones de actividad. Concretamente, con los datos monitorizados en la plataforma de gemelo digital el operador de la máquina pudo registrar y generar pautas de mantenimiento de la fresadora de forma interactiva sin interferir en las condiciones de trabajo.

#### 4.2.1. Nivel de interacción

En primer lugar se implementó la capa física de interacción utilizando un Gateway industrial portátil (modelo TWave T8-L mostrado en la **Figura 4a**). Esta solución incluye un dispositivo de hardware de adquisición personalizable para sensores comunes industriales, interfaces software con la nube y protocolos de comunicación industrial. Proporciona 12 canales de entrada de tipo BNC que aceptan medidas estáticas y dinámicas de diferentes tipos de sensores. Todas las señales capturadas se almacenan en una base de datos interna, donde se recogen series de datos temporales usadas para la monitorización de activos basada en la condición y la identificación de modos de fallo.

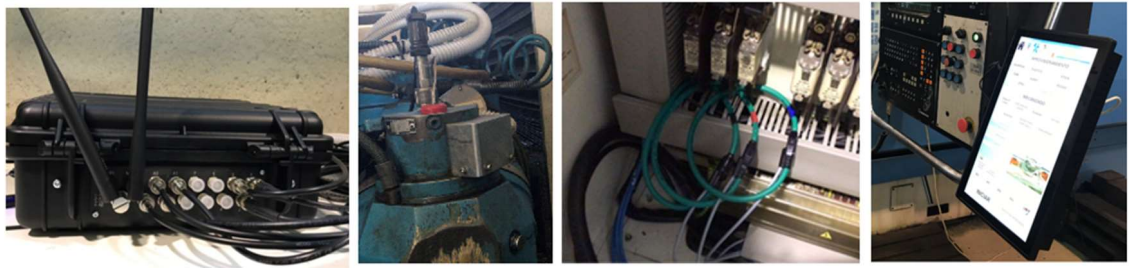

**Figura 4.** a) Gateway industrial TWave, b) Acelerómetros y sensor RDT, c) Transductor trifásico Rogowski, d) HMI panel PC táctil.

De acuerdo al enfoque no intrusivo de retro-actualización propuesto para no interferir con los sistemas instalados en la máquina, se utilizaron 3 tipos diferentes de sensores "plug & play" conectados por cable coaxial a los canales de entrada BNC del gateway industrial:

- Un sensor para detección de temperatura de resistencia magnética (RTD) tipo Pt100 adecuado para mediciones de alta temperatura en superficies ferrosas hasta un máximo de 300 ºC, con salida DC 4-20 mA.
- Dos acelerómetros PCB Piezotronics 603-Series con base magnética para montaje e instalación en superficies ferrosas.
- Un transductor de corriente AC trifásico para convertir la entrada de tensión a partir de tres lazos abiertos tipo Rogowski a salida DC 4-20 mA.

El sensor RDT y los dos acelerómetros se colocaron con la base magnética en el cabezal del mandrino de la fresadora, como se muestra en la **Figura 4b**. El transductor de corriente tipo Rogowski se instaló en el cuadro eléctrico de la fresadora colocando cada uno de los lazos en las líneas de corriente trifásica, como se muestra en la **Figura 4c**. De esta forma, todas las entradas fueron registradas para medir datos en tiempo real de la temperatura, vibración y consumo de la fresadora NC CF-20. Adicionalmente, para facilitar la monitorización de los registros tomados a pie de máquina y la interacción con el operador durante su funcionamiento, se instaló un equipo PC táctil como dispositivo HMI a la derecha del soporte del panel CNC, como se muestra en la **Figura 4d**. El gateway industrial cuenta además con un conector Ethernet para comunicación local con diferentes protocolos, entre ellos Modbus-TCP y HTTP, y de una conexión móvil 3G/4G para salida a Internet, lo que permite conectar el almacenamiento interno con la plataforma remota de gemelo digital en la nube.

### 4.2.2. Nivel de comprensión

En segundo lugar, se implementó la plataforma de aplicación y de servicio en la nube para construir el gemelo digital de la fresadora con diferentes modelos de datos. La representación virtual a través de la monitorización inteligente de cada uno de los subsistemas estudiados (con valores de consumo trifásico, temperatura y aceleración), se muestra en la **Figura 5**. Durante el estudio se registraron los arranques de la fresadora al principio de la jornada laboral, con el objetivo de extraer un conjunto de patrones asociados y modelar así su comportamiento. Para ello fue de mucha utilidad la experiencia aportada por el operador de la máquina. La ventaja de este enfoque es la facilidad para disponer en muy poco tiempo de series de datos temporales de funcionamiento de la fresadora que, en condiciones normales, cuentan con valores que se repiten cada día o por el contrario presentan una anomalía.

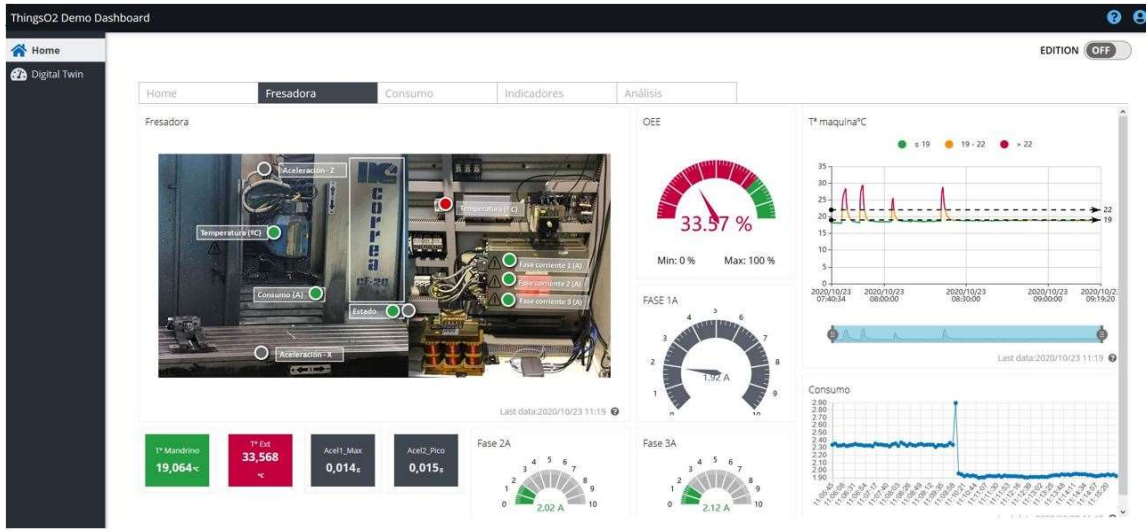

**Figura 5. Gemelo digital de la fresadora alojado en la plataforma en la nube**

Esta plataforma en la nube funciona en modo cliente-servidor, lo que permite la comunicación bidireccional con el Gateway industrial y la conexión con aplicaciones software en los dispositivos HMI. De esta manera se implementa la interconexión entre los valores físicos y los modelos de datos monitorizados. La retroalimentación del operador de la fresadora se obtiene a través del dispositivo HMI instalado a la derecha del panel CNC como se observa en la **Figura 6**. Consiste en un monitor táctil industrial que incluye una interfaz gráfica de aplicación software donde el trabajador registra las órdenes de trabajo en la fresadora. Las órdenes están formadas por el tipo de material, la operación de mecanizado, la herramienta utilizada y una codificación del plano CAD asociado.

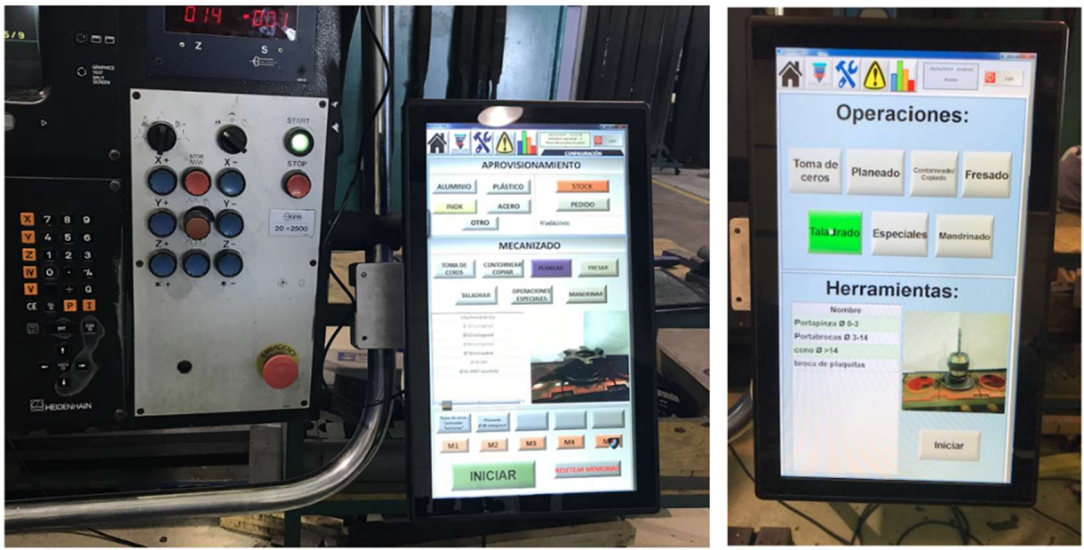

**Figura 6. Dispositivo HMI para interacción con el operador de la fresadora**

Además, el software instalado en el dispositivo HMI consulta los valores de aceleración y consumo de la fresadora para lanzar una ventana automática de cambio de operación/fin de trabajo cuando detecta inactividad. Esto permite registrar la duración de las operaciones que quedan etiquetadas en series de datos temporales. También se incluye un registro de incidencias para que el operador pueda registrar eventos, alarmas o aspectos que ayuden a entender anomalías encontradas durante el funcionamiento de la fresadora.

### 4.2.3. Nivel de aprendizaje

El último nivel implementado aborda un proceso de generación de conocimiento adaptativo con modelos de aprendizaje automático asociados a patrones de funcionamiento de la fresadora. Se apoya en la información registrada el cuadro de mando HMI y en una aplicación de realidad aumentada programada en Android que permite la comunicación interactiva del gemelo digital con el trabajador a través de una tableta Samsung Galaxy III y un código QR ubicado en diferentes zonas de la fresadora (ver **Figura 7**).

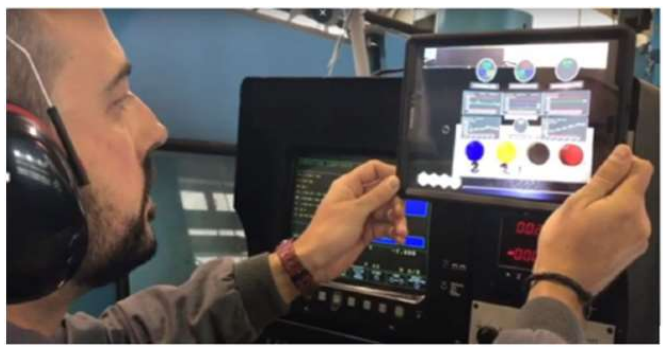

**Figura 7. Aplicación de realidad aumentada para monitorización interactiva**

Entre diciembre de 2019 y marzo de 2020 se monitorizó el arranque de la fresadora para recopilar datos diarios sobre múltiples arranques, proporcionando patrones de comportamiento repetitivos. Se extrajeron series de datos temporales del consumo de corriente a intervalos de un segundo, como se muestra en la **Tabla 1**.

| fecha | hora | Fase 1 (A) | Fase 2 (A) | Fase 3 (A) |
|---|---|---|---|---|
| 13/01/2020 | 8:09:50 | 0.68 | 0.09 | 0.58 |
| 13/01/2020 | 8:09:51 | 0.68 | 0.09 | 0.58 |
| 13/01/2020 | 8:09:52 | 2.15 | 1.75 | 2.63 |
| 13/01/2020 | 8:09:53 | 2.15 | 1.75 | 2.63 |
| 13/01/2020 | 8:09:54 | 2.15 | 1.75 | 2.63 |
| 13/01/2020 | 8:09:55 | 3.61 | 3.51 | 3.71 |
| 13/01/2020 | 8:09:56 | 3.81 | 3.51 | 3.81 |
| 13/01/2020 | 8:09:57 | 3.81 | 3.51 | 3.81 |
| 13/01/2020 | 8:09:58 | 3.51 | 3.61 | 3.81 |
| 13/01/2020 | 8:09:59 | 3.51 | 3.61 | 3.81 |

**Tabla 1. Aplicación de realidad aumentada para monitorización interactiva**

La detección y el proceso de aprendizaje se realizaron al principio de la jornada de trabajo. Para facilitar el proceso de modelado se hicieron grabaciones en la propia fresadora con un registro de marca de tiempo asociado al transductor de corriente trifásico. De esta forma, con toda la información recogida en el gemelo digital a partir del consumo de corriente, se pudieron observar 4 fases diferenciadas durante el arranque (ver figura **Figura 8a**): (i) encendido, (ii) carga del programa del PLC, (iii) arranque del motor, y (iv) calentamiento.

Como el operador debe realizar operaciones manuales sobre la fresadora en el arranque, éstas dependen de momentos variables. Por ejemplo, cuando se libera el botón de parada de emergencia para su rearme, previo al arranque del motor. Para disponer de conjuntos de datos comparables, se fijó el tiempo de duración de la fase 2 en 95 segundos interpolando con valores medios sin impactar en las medidas. En concreto, el proceso de modelado se realizó con un conjunto de datos correspondiente a 19 días entre los meses de enero y febrero de 2020. Además, se contó con la ayuda del trabajador para revisar los datos en el panel HMI de los diferentes arranques monitorizados. La representación y extracción de patrones sobre las muestras se realizó con utilidades de código abierto basadas en Python 3.x, como Numpy, Pandas, Matplotlib y Scipy.

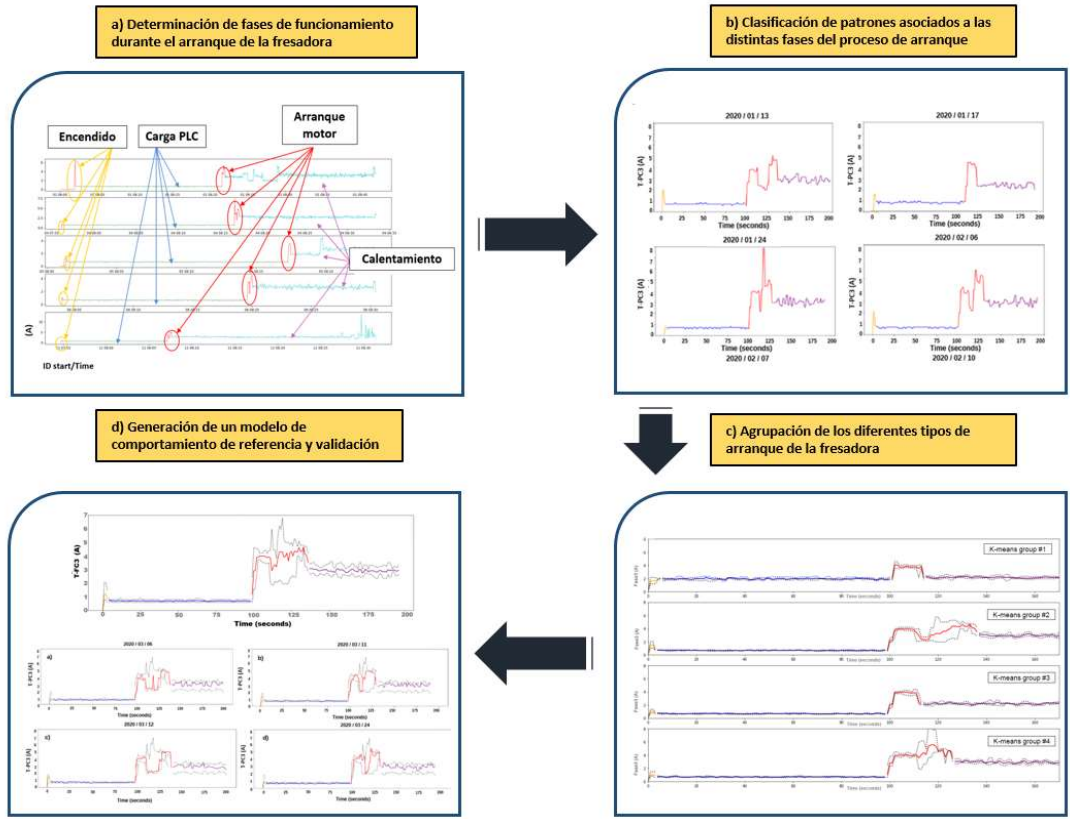

**Figura 8.** Varios pasos del proceso de generación de conocimiento para el arranque de la fresadora

Para minimizar el conjunto de muestras etiquetadas (ver **Figura 8b**), se aplicó el método de aprendizaje no supervisado k-means [30] de la librería gratuita de machine learning Scikit-learn basada en Python, que organiza de forma automática los patrones de arranque analizados de acuerdo a diferentes grupos o clusters. También se utilizó el método Elbow, llamado "del codo", para determinar el número óptimo de grupos entre las 19 muestras. En este caso el número fue igual a 4, como se muestra en la **Tabla 2**.

| fecha | grupo k-means | fecha | grupo k-means |
|---|---|---|---|
| 13/01/2020 | 2 | 03/02/2020 | 2 |
| 16/01/2020 | 1 | 04/02/2020 | 2 |
| 17/01/2020 | 3 | 05/02/2020 | 1 |
| 21/01/2020 | 3 | 06/02/2020 | 2 |
| 22/01/2020 | 1 | 07/02/2020 | 1 |
| 24/01/2020 | 4 | 10/02/2020 | 1 |
| 27/01/2020 | 3 | 11/02/2020 | 4 |
| 29/01/2020 | 4 | 13/02/2020 | 3 |
| 30/01/2020 | 4 | 14/02/2020 | 3 |
| 31/01/2020 | 2 | | |

**Tabla 2.** Grupos de clúster obtenidos durante en análisis del arranque de la fresadora

A continuación, se procesaron los valores mínimos, máximos y medios de todas las muestras de datos en cada conjunto de clusters para generar los umbrales límite en cada uno de los cuadro modelos (ver **Figura 8c**). Con el fin de generar un modelo único como referencia del proceso global de arranque de la fresadora, los cuatro modelos de partida fueron analizados por un operador experto. De esta forma, los valores límite admisibles para cada modelo fueron ajustados de forma supervisada de acuerdo al comportamiento aprendido durante las observaciones en tiempo real realizadas en tiempo real con el dispositivo HMI (Tablet Android). Por tanto, para la detección de los arranques anómalos de la fresadora se propuso como entrada al gemelo digital el modelo de referencia obtenido como la composición de los patrones comunes etiquetados por el operador (ver **Figura 8d**).

El modelo de comportamiento generado con el gemelo digital se sometió posteriormente a un conjunto de arranques de prueba durante el mes de marzo de 2020 para su validación (6 de marzo, 11 de marzo, 12 de marzo y 24 de marzo), considerándose adecuado el enfoque para el diagnóstico de anomalías que dependen de la condición de un activo, en este caso del consumo de corriente de la fresadora. Además, durante las pruebas con la aplicación de realidad aumentada, se pudo representar al mismo tiempo la vista física y la vista digital, comparando los valores medidos en tiempo real con la condición del activo durante las fases de arranque (ver ejemplo de alerta de corriente en la **Figura 9)**.

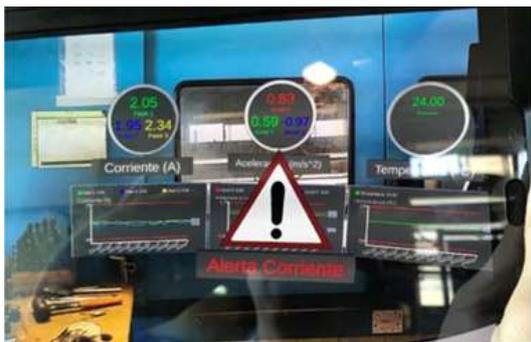

**Figura 9. Registro de alertas en tiempo real a través de realidad aumentada**

## 5. CONCLUSIONES

Durante décadas la industria de fabricación tradicional ha confiado a trabajadores expertos el control manual de la supervisión de sistemas y procesos. Es por esta razón que la convergencia entre los sistemas de fabricación heredados y las nuevas tecnologías de la información, mientras se incorporan los sistemas ciberfísicos, la inteligencia artificial y los gemelos digitales, es uno de los desafíos pendientes en las PYMEs. Sin embargo, la investigación de técnicas de retro-actualización adaptativas con bajo coste está facilitando la adopción de la convergencia del mundo físico con el digital. Se presenta por tanto la posibilidad de desplegar gemelos digitales de forma no intrusiva para la gestión de diferentes tipos de ecosistemas industriales en PYMEs. Al mismo tiempo aparecen oportunidades para solventar las limitaciones de conocimiento digital de los trabajadores, heredadas de los procesos de fabricación tradicionales.

El proyecto iDIGIT4L aporta una metodología de aprendizaje hombre-máquina que proporciona una interacción bidireccional y adaptativa en un escenario de fabricación tradicional. Se apoya en la implementación de un gemelo digital integrado de forma no intrusiva, caracterizando una vieja fresadora industrial para el aprendizaje con modelos de conocimiento que cuenta con la experiencia de trabajadores expertos. El uso de procedimientos aumentados y contenidos digitales para la interacción hombre-máquina ha resultado de gran utilidad para ahorrar tiempo y mejorar el rendimiento a través de un diagnóstico avanzado con la monitorización de indicadores en tiempo real. Como resultado se han podido actualizar al mismo tiempo las funcionalidades del sistema industrial y las habilidades digitales de los trabajadores. No obstante, se observa de manera muy preocupante que la resistencia al cambio en entornos tradicionales supone un obstáculo para la innovación. En particular cuando las nuevas tecnologías son probadas para su introducción en PYMEs. En futuras líneas de trabajo se debe abordar un proceso de aprendizaje que incorpore a todo tipo de trabajadores para romper las barreras de introducción de las nuevas tecnologías digitales. Es un hecho que la pérdida de puestos de trabajo va a afectar principalmente a los trabajadores de mayor edad y por ende en riesgo de exclusión de capacidades digitales, ya sea por una falta de conocimiento o por la falta de confianza para adaptarse a los nuevos requisitos de las plantas de fabricación del futuro.